\documentclass[prl,showpacs,amsfonts,amsmath,twocolumn,superscriptaddress,a4paper]{revtex4-1}

\usepackage{graphicx}
\usepackage{amsmath}
\usepackage{natbib}

\begin{document}

\title{Exploring the Effect of Noise on the Berry Phase}

\author{S.~Berger}
\email[]{sberger@phys.ethz.ch}
\affiliation{Department of Physics, ETH Zurich, CH-8093 Zurich, Switzerland}
\author{M.~Pechal}
\affiliation{Department of Physics, ETH Zurich, CH-8093 Zurich, Switzerland}
\author{A.~A.~Abdumalikov, Jr.}
\affiliation{Department of Physics, ETH Zurich, CH-8093 Zurich, Switzerland}
\author{C.~Eichler}
\affiliation{Department of Physics, ETH Zurich, CH-8093 Zurich, Switzerland}
\author{L.~Steffen}
\affiliation{Department of Physics, ETH Zurich, CH-8093 Zurich, Switzerland}
\author{A.~Fedorov}
\altaffiliation[Current address: ]{School of Mathematics \& Physics, The University of Queensland, Brisbane QLD 4072, Australia}
\author{A.~Wallraff}
\affiliation{Department of Physics, ETH Zurich, CH-8093 Zurich, Switzerland}
\author{S.~Filipp}
\affiliation{Department of Physics, ETH Zurich, CH-8093 Zurich, Switzerland}


\def\1{\mathchoice{\rm 1\mskip-4.2mu l}{\rm 1\mskip-4.2mu l}{\rm 1\mskip-4.6mu l}{\rm 1\mskip-5.2mu l}}
\newcommand{\ket}[1]{|#1\rangle}
\newcommand{\bra}[1]{\langle #1|}
\newcommand{\eval}[1]{\langle #1\rangle}
\newcommand{\braket}[2]{\langle #1|#2\rangle}
\newcommand{\ketbra}[2]{|#1\rangle\langle#2|}
\newcommand{\opelem}[3]{\langle #1|#2|#3\rangle}
\newcommand{\projection}[1]{|#1\rangle\langle#1|}
\newcommand{\scalar}[1]{\langle #1|#1\rangle}
\newcommand{\op}[1]{\hat{#1}}
\newcommand{\vect}[1]{\boldsymbol{#1}}
\newcommand{\id}{\text{id}}
\newcommand{\red}[1]{\textcolor{red}{#1} }

\newcommand{\sx}{\ensuremath{X}}
\newcommand{\sy}{\ensuremath{Y}}
\newcommand{\sz}{\ensuremath{Z}}

\newcommand{\rad}{\ensuremath{\, \mathrm{rad}}}
\newcommand{\khz}{\ensuremath{\, \mathrm{kHz}}}
\newcommand{\mhz}{\ensuremath{\, \mathrm{MHz}}}
\newcommand{\ghz}{\ensuremath{\, \mathrm{GHz}}}
\newcommand{\dbm}{\ensuremath{\, \mathrm{dBm}}}
\newcommand{\us}{\ensuremath{\, \mathrm{\mu s}}}
\newcommand{\ns}{\ensuremath{\, \mathrm{ns}}}
\newcommand{\mk}{\ensuremath{\, \mathrm{mK}}}
\newcommand{\um}{\ensuremath{\, \mathrm{\mu m}}}
\newcommand{\nm}{\ensuremath{\, \mathrm{nm}}}
\newcommand{\vpp}{\ensuremath{\, \mathrm{V_{pp}}}}

\date{\today}

\begin{abstract}
We experimentally investigate the effects of noise on the adiabatic and cyclic geometric phase, also termed Berry phase.
By introducing artificial fluctuations in the path of the control field, we measure the geometric contribution to dephasing of an effective two-level system for a variety of noise powers and different paths.
Our results, measured using a microwave-driven superconducting qubit, clearly show that only fluctuations which distort the path lead to geometric dephasing.
In a direct comparison with the dynamic phase, which is path-independent, we observe that the Berry phase is less affected by noise-induced dephasing.
This observation directly points towards the potential of geometric phases for quantum gates or metrological applications.
\end{abstract}

\maketitle

Noise is ubiquitous in physical systems---be it thermal noise in electrical circuits~\cite{Robinson1974}, electronic shot noise in mesoscopic conductors~\cite{Blanter1999}, vacuum noise of radiation fields~\cite{Glauber1963}, or low-frequency \mbox{(1/f-)} noise in solid state systems~\cite{Dutta1981,Bylander2011}.
It prevents quantum coherence to persist on long time scales and hinders the development of a large-scale quantum computer~\cite{Schlosshauer2007,Joos2003}.
Significant effort has thus been put into concepts and methods to control and maintain fragile quantum superposition states~\cite{Ladd2010}. The geometric phase is a promising building block for noise-resilient quantum operations~\cite{Sjoqvist2008} and its properties in open quantum systems have been actively investigated in theory~\cite{Blais2003,De2003,Whitney2005,Whitney2003,Carollo2003a,Solinas2010,Villar2011,Solinas2012}. There are, however, only a few experiments studying the contribution to dephasing stemming from the Berry phase~\cite{Filipp2009a,Cucchietti2010,Wu2013a}.

\begin{figure*}
  \centering
 \includegraphics[width=160mm]{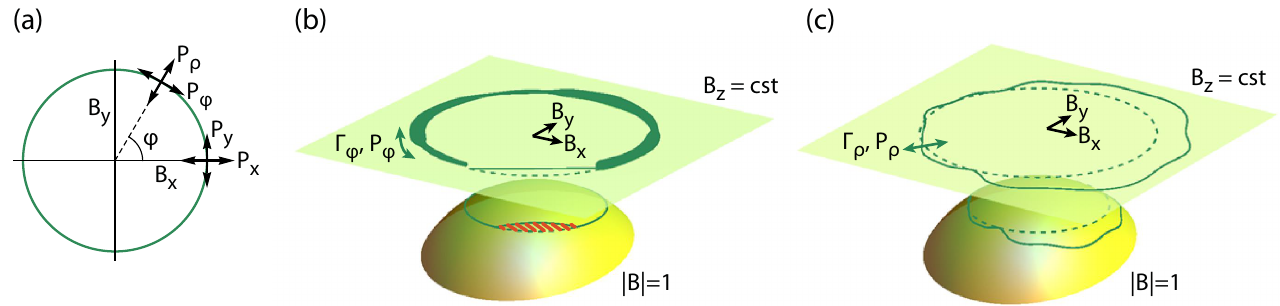}
  \caption{
(a)
The path of the effective field (green line) describes a circle in the $B_x$-$B_y$-plane at constant $B_z$. Noise in $x$ and $y$ directions with noise powers $P_x$ and $P_y$ can be decomposed into noise in $\rho$ and $\varphi$ directions with noise powers $P_\rho$ and $P_\varphi$.
(b,c)
The path of the effective field without noise (dashed green lines lying in the plane with constant $B_z$) is drawn alongside the same path exposed to two kinds of noise (solid green lines): angular noise in (b), where the velocity of precession is proportional to line thickness, and radial noise in (c). The projection of the paths on the unit sphere $|\mathbf{B}|=1$ is also shown. In (b), the difference in solid angle due to non-cyclic evolution is highlighted in red.
  }
  \label{fig:1}
\end{figure*}

In this Rapid Communication, we study the physics of a two-level system, a qubit, in an effective field $\mathbf{B}$, described by the Hamiltonian
\begin{equation}
\label{eq:ham}
H=\hbar\boldsymbol\sigma\cdot\mathbf{B}/2,
\end{equation}
where $\boldsymbol\sigma=(\sx,\sy,\sz)$ are the Pauli matrices, and $\mathbf{B}=(B_x,B_y,B_z)$ is given in units of angular frequency. If the field is adiabatically and cyclically varied in time, the ground $\ket{0}$ and excited state $\ket{1}$ of the two-level system acquire a geometric phase $\gamma_0=\pm A/2$, where $A$ is the solid angle (with respect to the origin $\mathbf{B}=0$) enclosed by the path traced out by $\mathbf{B}(t)$~\cite{Berry84}. This type of geometric phase is known as Berry phase. Here, we consider an effective field evolving along a circular path with radius $B_\rho=\sqrt{B_x^2+B_y^2}$ at constant $B_z$ and with precession period $\tau$ (Fig.~\ref{fig:1}). This path encloses a solid angle $A=2\pi(1-\cos\vartheta)$, with the polar angle $\vartheta=\arctan(B_\rho/B_z)$.

In realistic situations, the field components fluctuate about their mean values and these fluctuations induce dephasing. Changes in field strength will cause dynamic dephasing, while modifications in solid angle will cause geometric dephasing. Clearly, noise directed in azimuthal direction [angular noise, Fig.~\ref{fig:1}(b)] does not modify the solid angle and thus, no geometric dephasing is expected. In contrast, noise directed in radial direction [radial noise,  Fig.~\ref{fig:1}(c)] will lead to geometric contributions to dephasing. By artificially adding noise in radial (or azimuthal) direction to the field in our experiment, we are thus able to maximize (or minimize) geometric dephasing and investigate its properties for different angles $\vartheta$ and noise powers.

To model realistic uncorrelated noise with a given bandwidth, we generate fluctuations conforming to Ornstein-Uhlenbeck processes, i.e.~stationary, gaussian and markovian noise processes with a lorentzian spectrum of bandwidth $\Gamma_i$ and noise power $P_i$ ($i=\rho,\varphi)$.
In the experiment, the precession frequency and the noise bandwidth are chosen to be  small compared to the amplitude $B=|\mathbf{B}|$ of the effective field, i.e.~$1/\tau,\Gamma_i\ll B$, to study adiabatic processes. In this case, we can derive the variance of the geometric phase from a perturbative treatment. To first order in the noise variations $\delta\varphi$ and $\delta\rho$, the deviation $\delta\gamma$ of the Berry phase is~\cite{De2003}
\begin{equation}
\label{eq:deviationG}
\delta\gamma = -\frac{\pi}{\tau}\int_0^\tau\sin\vartheta\delta\vartheta \mathrm{d}t.
\end{equation}
As the ensemble average 
of $\delta\gamma$ vanishes, the mean Berry phase is identical to $\gamma_0$. Expressing the effective field in cylindrical coordinates, $\mathbf{B}=(B_\rho\cos\varphi,B_\rho\sin\varphi,B_z)$, the variations in the polar angle can be written as $\delta\vartheta=(\cos\vartheta/B)\delta\rho$, and the Berry phase is found to have a gaussian distribution with variance
\begin{equation}
\label{eq:varianceG}
\sigma_\gamma^2 = 2P_\rho \left(\frac{\pi\cos\vartheta\sin\vartheta}{B\tau}\right)^2
      \frac{\Gamma_\rho \tau-1+e^{-\Gamma_\rho \tau}}{\Gamma_\rho^2}.
\end{equation}
As expected, to first order only variations $\delta\rho$ in radial direction contribute to $\sigma_\gamma^2$.

Geometric phases have been observed in a variety of superconducting systems~\cite{Leek2007,Mottonen2008,Neeley2009,Berger2012}. Here, 
we use the two lowest energy levels of a superconducting artificial atom of the transmon type~\cite{Koch2007} embedded in a transmission line resonator---an architecture known as circuit quantum electrodynamics~\cite{Blais2004,Wallraff2004a}. Note, however, that our findings are independent of the specific implementation, and apply to any system in which Berry phases can be observed.
The qubit is manipulated using microwave fields applied via a capacitively coupled charge bias line. Using spectroscopic measurements, we have determined the maximum Josephson energy $E_{J,\mathrm{max}}/h=11.4\ghz$, the charging energy $E_C/h=0.26\ghz$
and the coupling strength $g/2\pi=360\mhz$ of the qubit to the resonator.
The experiments are performed at a qubit transition frequency $\omega_{01}/2\pi=4.68\ghz$, with an energy relaxation time $T_1=2.65\us$, a phase coherence time $T_2=1.35\us$ and a spin-echo phase coherence time $T_2^{\mathrm{echo}}=2.15\us$. The sample is operated in a dilution refrigerator at a base temperature of $20\mk$.
In the dispersive regime, when $\omega_{01}$ is far detuned from the resonator mode, the Hamiltonian of the driven system is~\cite{Leek2007}
\begin{equation}
\label{eq:ham_eff}
H_{\mathrm{eff}} = \hbar\left(\sx\Omega\cos\varphi + \sy\Omega\sin\varphi + \sz\Delta \right)/2
\end{equation}
in a reference frame which rotates at the drive frequency $\omega_d$.
This Hamiltonian is identical to the one in equation (\ref{eq:ham}) with an effective field $\mathbf{B}=(\Omega\cos\varphi,\Omega\sin\varphi,\Delta)$. It is determined by amplitude $\Omega$, phase angle $\varphi$ and detuning $\Delta=\omega_{01}-\omega_d$ of the drive.

\begin{figure}
  \centering
 \includegraphics[width=86mm]{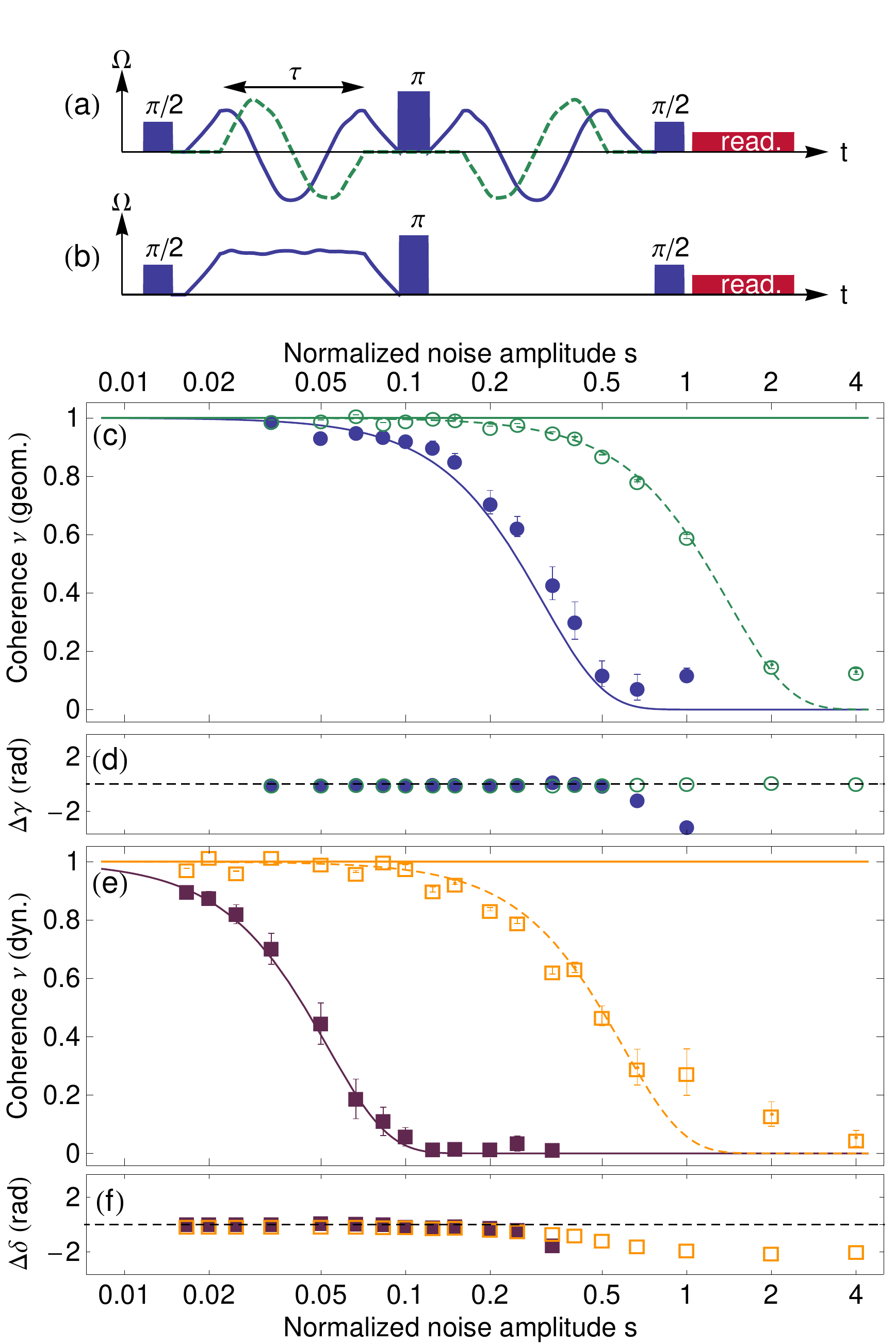}
  \caption{
(a,b)
Sketches of the pulse schemes used to measure (a) Berry and (b) dynamic phases with radial noise. Pulses applied along the $x$ and $y$ quadratures are shown in blue and green, respectively. The readout pulse (red, see text) concludes the sequence after $t\approx 400$ ns.
(c,d)
Experimentally measured coherence $\nu$  of the Berry phase and phase difference $\Delta\gamma=\gamma-\gamma_0$ as a function of normalized noise amplitude $s$ for radial noise (filled blue circles) and angular noise (open green circles), plotted on a logarithmic scale. For every value of $s$, 300 noise realizations were measured with noise bandwidth $\Gamma=10\mhz$ at solid angle $A=7\pi/16$ and evolution time $\tau=100\ns$. The continuous line is computed from Eq.~(\ref{eq:varianceG}). The dashed line is a fit to the function $\exp(-(4as)^2/2)$ with fitting parameter $a=0.25\pm0.01$.
(e,f)
Quantities analogous to panels (c,d) but for the dynamic phase, with $\Delta\delta=\delta-\delta_0$ and fitting parameter $a=0.60\pm0.03$.}
  \label{fig:3}
\end{figure}

A Ramsey-type interferometric sequence containing a spin-echo pulse to cancel the dynamic phase~\cite{Filipp2009a,Jones2000} is employed to measure the Berry phase acquired by the two-level system [Fig.~\ref{fig:3}(a)]. A series of resonant pulses (of frequency $\omega_{01}$) implement the spin-echo sequence, while  off-resonant pulses (of frequency $\omega_d=\omega_{01}-\Delta$) guide its state adiabatically along the paths sketched in Fig.~\ref{fig:1}(b,c).

All presented Berry phases are measured at a detuning $\Delta=-50\mhz$. %
The acquired Berry phase is varied from $0$ rad to $6.9$ rad by increasing the solid angle $A$ via the drive amplitude $\Omega$. The strength of the noise is quantified by the normalized noise amplitude $s_\rho=\sqrt{P_\rho}/B_\rho$ for radial noise and by $s_\varphi=\sqrt{P_\varphi}$ for angular noise. These definitions ensure that fluctuations in radial or azimuthal directions have identical amplitudes if $s_\rho=s_\varphi$.

The phases with noise are obtained by repeating the experiment with different noise patterns. Identical noise patterns are used before and after the spin echo pulse to ensure cancellation of the dynamical phase. The pulse sequences, consisting of two intermediate-frequency quadratures $x$ and $y$, are numerically created: noise conforming to an Ornstein-Uhlenbeck process is generated and added to the pulses describing the noiseless evolution of the field. An arbitrary waveform generator synthesizes these quadratures, which are upconverted to a microwave-frequency signal using an in-phase/quadrature mixer. After the manipulation sequence, the state of the qubit is determined in a dispersive readout~\cite{Bianchetti2009} through the resonator and reconstructed using state tomography~\cite{Paris2004}. To overcome noise in the detection, each individual noise realization is measured $10^6$ times.

\begin{figure}
  \centering
 \includegraphics[width=86mm]{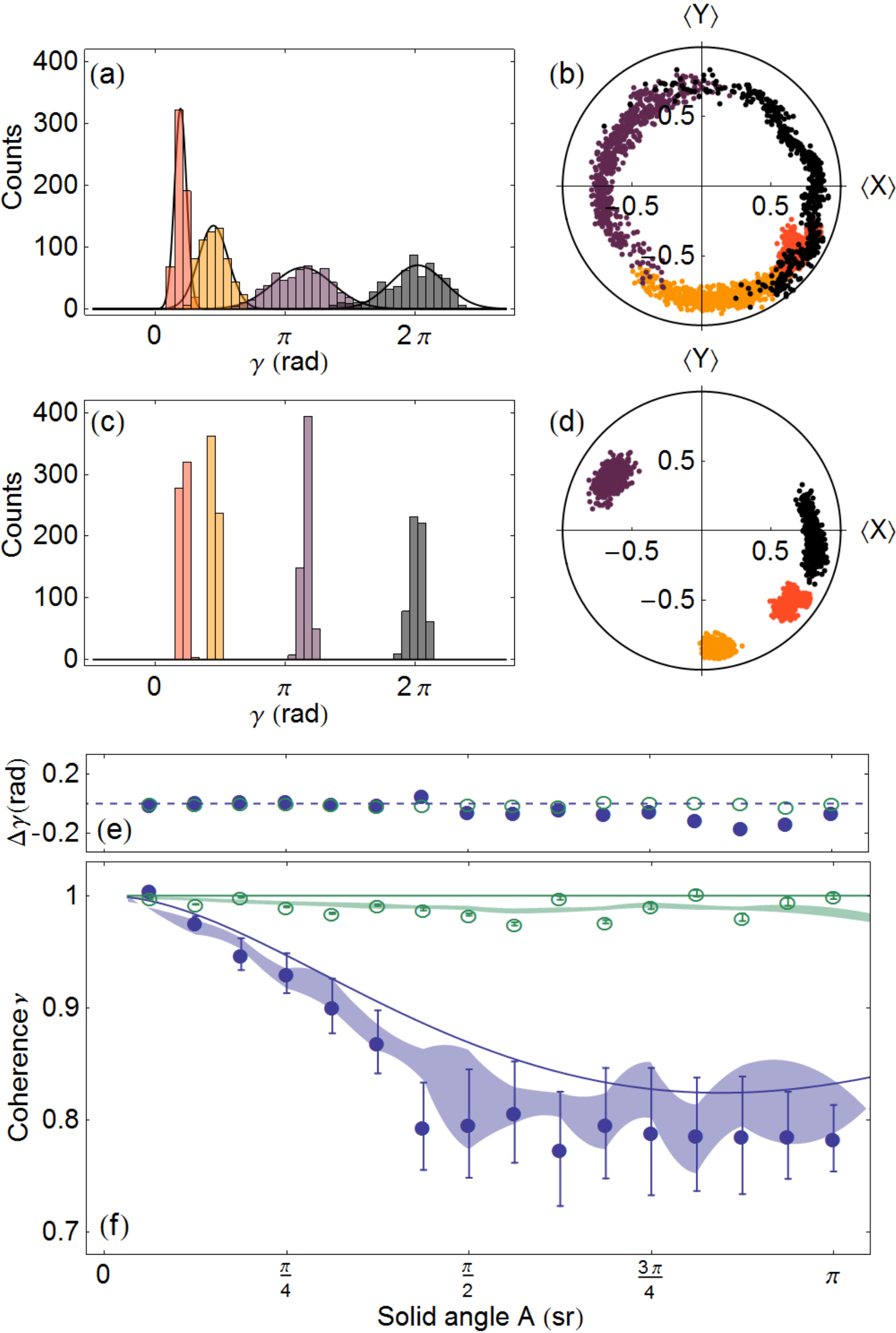}
  \caption{
(a)
Histograms of Berry phases and (b) measured expectation values $\eval{\sx}$, $\eval{\sy}$ of 600 realizations of radial noise for each solid angle $A=\pi/16,3\pi/16,8\pi/16$ and $15\pi/16$ (indicated in red, orange, purple and black). Fits of a gaussian to the measured histograms are also shown in (a). The circle in (b) indicates unit coherence.
(c,d)
Measurements analogous to panels (a,b) for angular noise.
(e,f)
Coherence $\nu$ and phase difference $\Delta\gamma$ as a function of solid angle $A$ for radial noise (filled blue circles) and angular noise (open green circles). The experimental data points are shown alongside the theory curve (solid lines) and the results from numerical simulations (the shaded area indicates the standard deviation about the mean). Data in panels (a-f) is recorded at fixed noise bandwidths $\Gamma_i=10\mhz$, normalized noise amplitudes $s_i=1/15$ and evolution time $\tau=100\ns$.}
  \label{fig:2}
\end{figure}

Histograms of the measured Berry phases for four solid angles are shown in Fig.~\ref{fig:2}(a-d). For radial noise, the Berry phases of the individual noise realizations have---as discussed above---a gaussian distribution with a mean equal to the Berry phase $\gamma_0$ without noise. For angular noise, we observe that the widths of the phase distributions are, as expected, almost zero.
The expectation values of the Bloch-vector components $\eval{\sx}$ and $\eval{\sy}$ for individual noise realizations are distributed on the equatorial plane of the Bloch sphere (Fig.~\ref{fig:2}b,d), reflecting the spread of the measured phases. They lie on a circle with radius $\nu_0\approx0.80<1$, which is a result of the intrinsic noise present in the system.
%

Distributions akin to those shown in Fig.~\ref{fig:2}(b,d) are used to compute the coherence $\nu=\sqrt{\eval{\sx}^2+\eval{\sy}^2}=e^{-(4\sigma_\gamma)^2/2}$ versus solid angle [Fig.~\ref{fig:2}(f)].
In this plot and all subsequent plots, the coherences are normalized to a measurement without added noise whereby the intrinsic noise is eliminated.
We observe that for radial noise the coherence decreases and then stabilizes as a function of solid angle, while it is approximately unity for angular noise. This is an immediate consequence of the nature of the Berry phase: radial noise modifies the solid angle $A$ causing dephasing and a decrease in coherence. In contrast, angular noise hardly affects $A$.

For both kinds of noise, the difference $\Delta\gamma=\gamma-\gamma_0\lesssim0.2$ rad in the mean Berry phase with and without noise is very small [Fig.~\ref{fig:2}(e)]. The measured coherences agree well with equation~(\ref{eq:varianceG}) and numerical results obtained by solving the unitary dynamics of the Hamiltonian in equation~(\ref{eq:ham_eff}).
The measured Berry phase $\gamma_0$ itself (not shown) agrees well with the prediction for a transmon-type qubit~\cite{Berger2012}, with a discrepancy of $0.20$ rad across all solid angles for the data in Fig.~\ref{fig:2}(e,f).

To illustrate the effects of noise quantitatively, both
the Berry phase and the dynamic phase are measured for varying noise amplitudes $s$. For the Berry phase, we observe that the coherence follows the expected dependence $e^{-(4as)^2/2}$ for radial noise [Fig.~\ref{fig:3}(c,e)] and that angular noise has a lesser effect on the coherence than radial noise. For both types of noise, and for normalized noise amplitudes $\lesssim0.5$, the Berry phase with and without noise have the same value.

The coherence of  the dynamic phase $\delta$ can be computed perturbatively, in the same way as for the Berry phase. Using the deviation $\int_0^\tau\delta B\, \mathrm{d}t / \hbar = \int_0^\tau\sin\vartheta\, \delta\rho\, \mathrm{d}t / \hbar$
of the dynamic phase, one finds its mean $\delta$ and its variance
\begin{equation}
\label{eq:varianceD}
\sigma_\delta^2 = 2P_\rho (\sin\vartheta)^2
       \frac{\Gamma_\rho \tau-1+e^{-\Gamma_\rho \tau}}{\Gamma_\rho^2}.
\end{equation}
Only radial variations contribute to $\sigma_\delta^2$ and cause the dynamic phase to have a gaussian distribution around the noiseless dynamic phase $\delta_0$. Noise in azimuthal direction does not change the magnitude of the field and hence does not cause fluctuations in the dynamic phase.

The coherence of the dynamic phase was recorded using a spin-echo sequence containing a single off-resonant pulse (Fig.~\ref{fig:3}b), and therefore its variance was scaled by a factor to allow for direct comparison with the Berry phase. From Fig.~\ref{fig:3}(e), it is evident that the coherence of the dynamic phase starts decreasing at weaker noise amplitudes than the Berry phase, demonstrating the superior noise resilience of the Berry phase. It is also observed that the mean dynamic phase $\delta$ starts deviating from $\delta_0$ already at $s\approx0.2$. The measured coherences for both dynamic and Berry phase are in very good agreement with the predictions based on equations (\ref{eq:varianceG}) and (\ref{eq:varianceD}) for radial noise. For angular noise, fits to $e^{-(4as)^2/2}$ agree with the observed behaviour of the coherences. Indeed, while according to equations (\ref{eq:varianceG}) and (\ref{eq:varianceD}) the coherences are expected to be insensitive to angular noise to first order, non-adiabatic and higher-order effects \cite{Lupo2009} still affect the coherences.
In particular, the evolution of the field can be non-cyclic~\cite{Samuel1988}, which adds a small contribution to dephasing~\cite{De2003} [Fig.~\ref {fig:1}(b)].

\begin{figure}
  \centering
 \includegraphics[width=86mm]{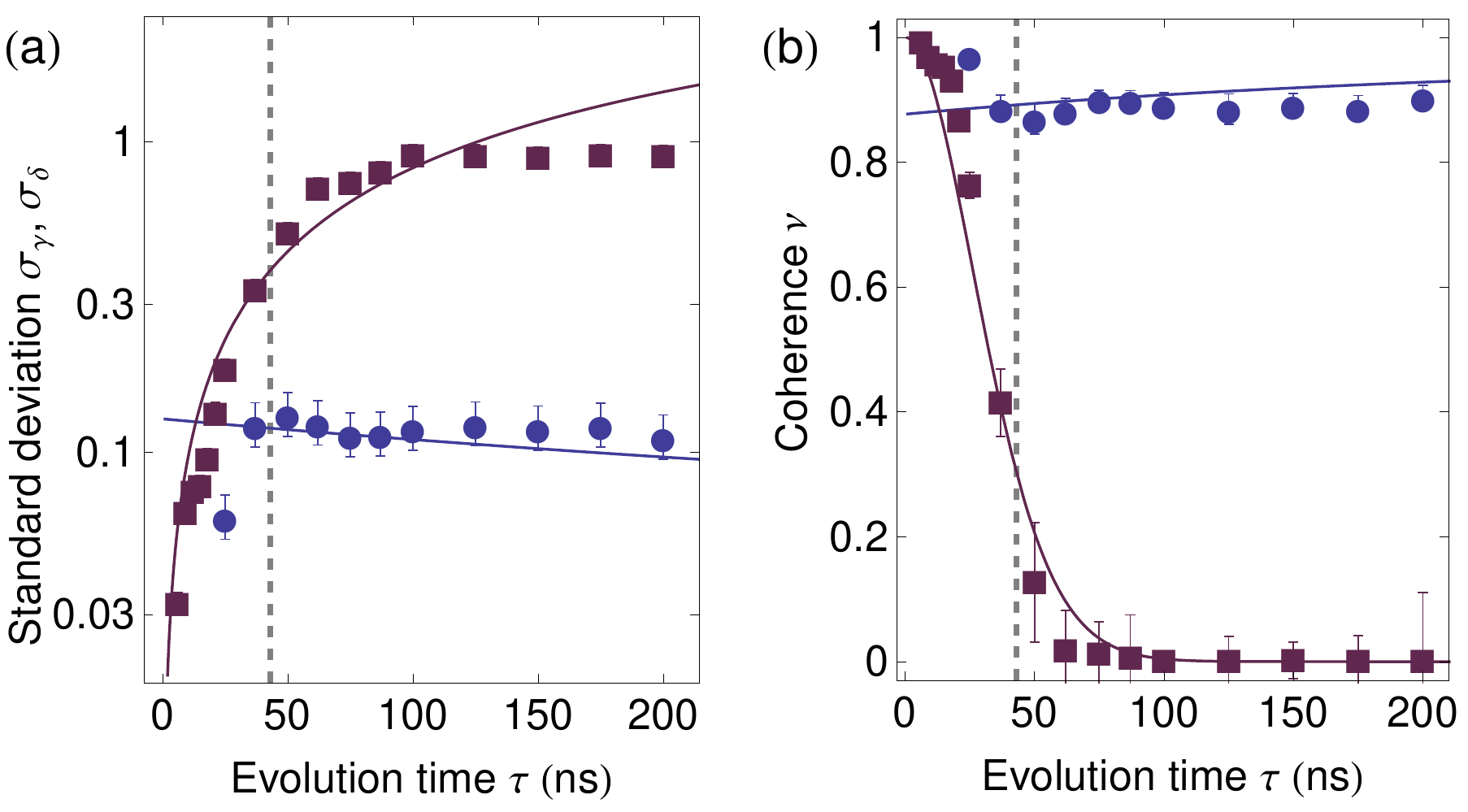}
  \caption{
(a)
Standard deviation $\sigma_\gamma$ of the Berry phase (blue circles) and $\sigma_\delta$ of the dynamic phase (purple squares) as a function of evolution time $\tau$, based on 300 noise realizations with $\Gamma=10\mhz$ and $s_{\rho}=1/15$. The solid lines result from calculations based on Eqs.~(\ref{eq:varianceG}) and (\ref{eq:varianceD}). The dashed grey line approximately separates the non-adiabatic from the adiabatic regime.
(b)
Coherence $\nu$ versus evolution time $\tau$ of the Berry phase (blue circles) and the dynamic phase (purple squares).
  }
  \label{fig:4}
\end{figure}

Finally, we directly compare the coherence of dynamic and Berry phases in the presence of radial noise. The Berry phase $\gamma$ is recorded at a solid angle $A=0.37\pi$, where the effect of noise on $\gamma$ is strongest. For long evolution times $\tau$, the Berry phase is more resilient against radial noise than the dynamic phase because its variance $\sigma_\gamma^2$ decreases with evolution time \cite{Filipp2009a}, whereas the variance of the dynamic phase $\sigma_\delta^2$ grows linearly in evolution time (cf.~equations~(\ref{eq:varianceG}) and (\ref{eq:varianceD}), as well as Fig.~\ref{fig:4}). Both phases have equal coherences when
$\sigma_\gamma^2=\sigma_\delta^2$, i.e.
\begin{equation}
\label{eq:threshold}
\tau=\pi\cos(\vartheta)/B,
\end{equation}
and the dynamic phase is more coherent than the Berry phase only for even shorter evolution times ($\tau<13$ ns according to equation~(\ref{eq:threshold}) and $\tau<20$ ns according to the experimental data in Fig.~\ref{fig:4}). Note that the variance of the dynamic phase is independent of the value of the dynamic phase, this is why it was recorded using the same drive amplitudes as for the Berry phase gates.
The data in Fig.~\ref{fig:4} agrees with calculations. The standard deviation $\sigma_\delta$ of the dynamic phase starts differing significantly from computed predictions at evolution times $\tau\gtrsim100\ns$, when the recorded phases are spread across $2\pi$ and their variance saturates.

In conclusion, we have demonstrated that the Berry phase is less affected by noise along the path in parameter space than by noise perpendicular to it. Given a system with known noise properties, this can potentially be exploited to realize noise-resilient geometric operations. Both kinds of noise leave the mean of the geometric phase unchanged. Shifts of the mean Berry phase are theoretically expected~\cite{Whitney2005}, but are beyond current experimental precision. We have also shown that the geometric phase is less affected by decoherence than the dynamic phase when evolving adiabatically (evolution times $\gtrsim1/B$). Our results beautifully exemplify fundamental properties of the geometric phase and serve as a stepping-stone for further investigations of geometric phases as a resource for quantum computation or for precision measurements \cite{Martin-Martinez2013,Rong2011,Pikovski2012,Mur-Petit2012}.

S.B. thanks A. Agazzi for help with numerical simulations. This work was supported by the Swiss National Science Foundation (SNF) and the EU project GEOMDISS.



%

\end{document}